\documentclass{article}%[9pt,twocolumn,twoside]
\newcommand*\chem[1]{\ensuremath{\mathrm{#1}}}
\usepackage{multicol}
\usepackage{graphicx}
\usepackage{color}
\usepackage{amsmath}
\usepackage{amssymb}
\usepackage{enumitem}
\usepackage[margin=0.6in]{geometry}
\usepackage{MnSymbol}
\usepackage{cite}
\usepackage{authblk}
\begin{document}
\title{Origin of third harmonic generation in plasmonic nanoantennas}

\author[1,2,*]{Antonino Cala' Lesina}
\author[1,2,3]{Pierre Berini}
\author[1,2,*]{Lora Ramunno}

\affil[1]{Department of Physics, University of Ottawa, Ottawa, Canada}
\affil[2]{Centre for Research in Photonics, University of Ottawa, Ottawa, Canada}
\affil[3]{School of Electrical Engineering and Computer Science, University of Ottawa, Ottawa, Canada}

\affil[*]{Corresponding author: antonino.calalesina@uottawa.ca, lora.ramunno@uottawa.ca}

%\dates{Compiled \today}

%\ociscodes{(190.0190) Nonlinear optics; (190.2620) Harmonic generation and mixing; (190.4350) Nonlinear optics at surfaces; (190.4400) Nonlinear optics, materials; (310.6628) Subwavelength structures, nanostructures; (290.0290) Scattering.}

%\doi{\url{http://dx.doi.org/10.1364/ao.XX.XXXXXX}}

\maketitle

\begin{abstract}
Plasmonic nanoantennas have been recently proposed to boost nonlinear optical processes. In a metal dipole nanoantenna with a dielectric nanoparticle placed in the gap, the linear field enhancement can be exploited to enhance third harmonic emission.
Since both metals and dielectrics exhibit nonlinearity, the nonlinear far-field contains contributions from each, and the impossibility of measuring these contributions separately has led to seemingly contradictory interpretations about the origin of the nonlinear emission. We determine that the origin of the third harmonic from metal-dielectric dipole nanoantennas depends on nanoantenna design, and in particular, the width. We find that the emission from gold dominates in thin threadlike nanoantennas, whereas the emission from the gap material dominates in wider nanoantennas. We also find that monopole nanoantennas perform better than dipoles having the same width, and due to their simplicity should be preferred in many applications.
\end{abstract}

%\setboolean{displaycopyright}{true}

%\thispagestyle{fancy}

%\ifthenelse{\boolean{shortarticle}}{\ifthenelse{\boolean{singlecolumn}}{\abscontentformatted}{\abscontent}}{}
%\begin{multicols}{2}
\section{Introduction}
Noble metal nanostructures have been explored in recent years for their ability to confine light over subwavelength volumes with inherent field enhancement of the incoming radiation. 
This is ascribed to plasmonic resonances induced in the nanostructures. Nonlinear optics in bulk media requires phase matching to maximize the nonlinear emission. Due to the field enhancement in the proximity of plasmonic nanoantennas, nonlinear optical processes can be enhanced at the nanoscale with no phase matching requirement \cite{Zayats2009}. The field enhancement takes place in gaps between nanostructures, near sharp corners, in film-coupled nanoparticles \cite{Ciraci2012a}, and inside the metal for nanoantenna sizes of the same order as the skin depth. The field enhancement has been exploited to enhance nonlinear optical processes, such as second harmonic generation (SHG) \cite{Chen2015b}, third harmonic generation (THG) \cite{Lassiter2014,Jin2016}, four-wave mixing (FWM) \cite{Genevet2010}, and difference frequency generation (DFG) \cite{CalaLesina2015}.

Hybrid dielectric/plasmonic nanostructures, such as a metallic dipole nanoantenna with a nonlinear dielectric nanoparticle in the gap, have been proposed to enhance nonlinear emission.
Since both metals and dielectrics exhibit nonlinearity, the nonlinear far-field contains contributions from each. The impossibility of measuring these contributions separately has led to controversial interpretations about the origin of the nonlinear signal. For this reason, though the hybrid dipole nanoantenna is well understood in the linear regime, in the nonlinear regime the origin of nonlinear emission is still under debate \cite{Utikal2011,Aouani2014,Metzger2014,Linnenbank2016}.
In \cite{Aouani2014} a dipole nanoantenna with an \chem{ITO} disk in the gap was reported, and the THG emission was attributed to \chem{ITO} due to the linear field enhancement in the gap.
In \cite{Metzger2014} a similar experiment was carried out and a doubled THG intensity was observed. The conclusion there was that the THG signal comes primarily from gold, and the enhanced nonlinear radiation was attributed to the change of the linear properties of the nanoantenna due to the presence of the dielectric in the gap.

In this paper, we use a full numerical approach to explore the design space of the hybrid dipole nanoantenna and illustrate regimes where THG is dominated by the gap dielectric, by the metal or by both. 
A detailed analysis is required to quantify how each part of the system contributes to the third harmonic emission. We consider THG because it requires less complex structures, {\it e.g.}, a monopole nanoantenna, which is not suitable for SHG \cite{Ciraci2012,Li2015}. Furthermore, the third order nonlinear susceptibility can be derived from the linear properties of the medium based on Miller's rule, whereas the application of this rule is not straightforward for SHG in nanostructures and metamaterials \cite{OBrien2015}. The design space of a metal monopole nanoantenna is also explored to show when this simpler structure should be preferred for THG.

\section{Results}

\begin{figure}[htbp]
\centering
\includegraphics[width=0.45\textwidth]{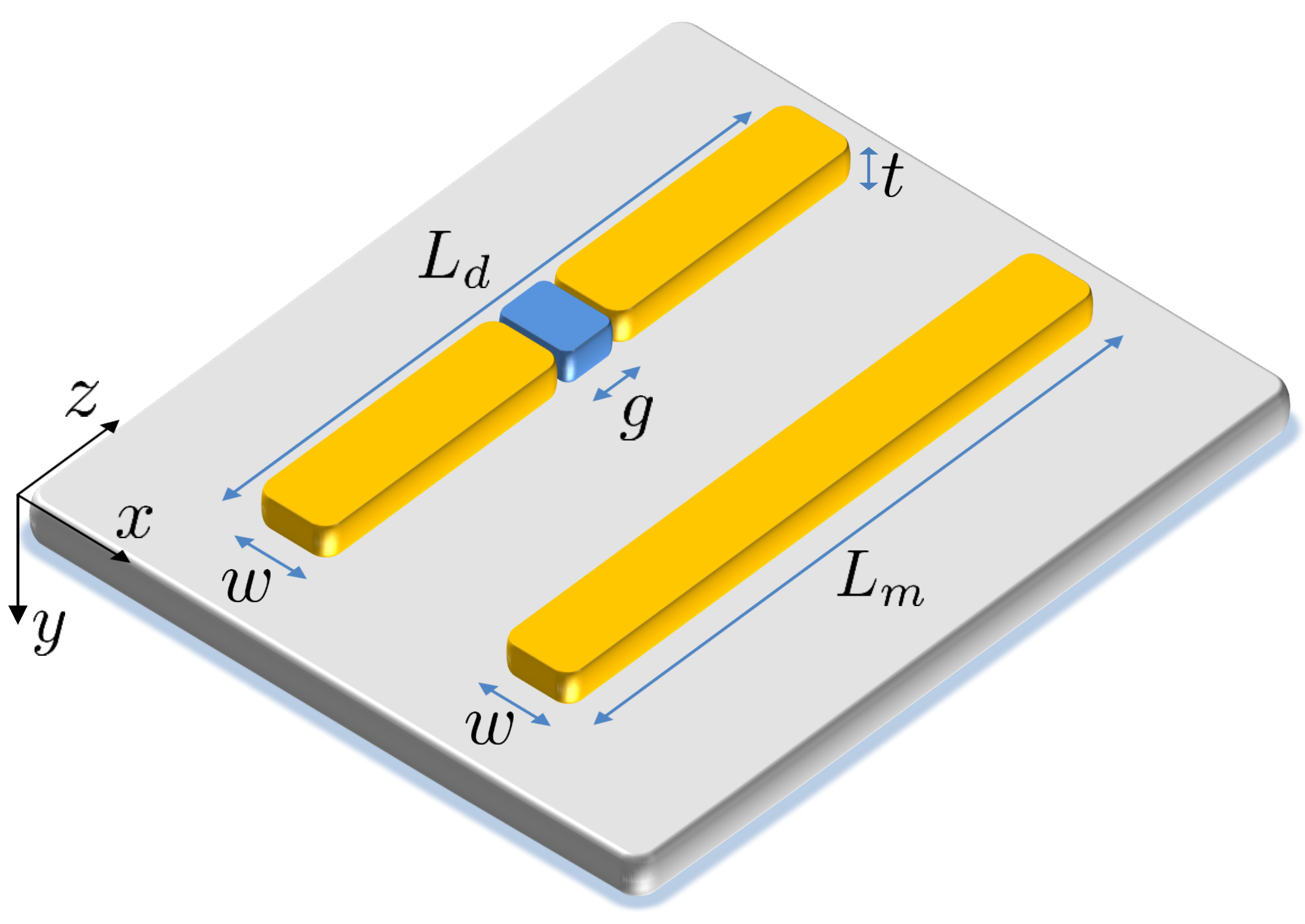}
\caption{Sketch of dipole and monopole nanoantennas.}
\label{fig1}
\end{figure} 

We use the finite-difference time-domain (FDTD) method \cite{CalaLesina2015a} to simulate in 3D gold nanostructures, {\it i.e.}, dipole and monopole nanoantennas, on a semi-infinite \chem{SiO_2} substrate (Fig. \ref{fig1}) exposed to air. 
The nanoantennas are oriented along the $z$-direction on an $xz$-plane. A uniform space-step of 1 nm is used. % to discretize the simulation domain.
A $z$-polarized and $y$-propagating plane wave is introduced from air by the total-field/scattered-field (TF/SF) technique. The monochromatic excitation at $\omega_0$ ($\lambda_0=900$ nm) is implemented by a sinusoidal signal modulated by a Gaussian pulse. 
To monitor the far-field of the THG emission, we use a large simulation domain of (800 $nm)^3$. The analytical Kirchhoff near-to-far field transformation cannot be applied, because the nanoantenna is not embedded in a homogeneous medium.
The dipole nanoantenna has variable length within the range $255\leq L_d\leq 324$ nm and a gap $g=\{10,20,30\}$ nm. The monopole nanoantenna has variable length in the range $174\leq L_m\leq 198$ nm. 
The thickness of all the nanostructures is $t=50$ nm, and the width $w = \{30,50,70\}$ nm. 
The gap of the dipole nanoantenna is filled by a nanoparticle of length $g$ and width $w$.
%In order to compare the nonlinear emission,
The lengths $L_d$ and $L_m$ were adjusted such that each nanoantenna is resonant at $\lambda_0$. The lengths were obtained by linear interpolation, having verified that in the examined wavelength range the resonance wavelength shifts linearly with the nanoantenna length, as reported in \cite{Mousavi2012}.
The corners and edges of the metal nanostructures are rounded by $r=10$ nm curvature radius to reduce numerical artifacts and to neglect nonlocal effects. 
Linear dispersion is modelled by the auxiliary differential equation (ADE) technique: gold by the Drude+2CP model \cite{Vial2011}, \chem{SiO_2} by the Lorentz model using experimental data in \cite{Malitson1965}, and \chem{ITO} by the Drude model \cite{Luk2015}. 
The THG process is modelled by adding an isotropic instantaneous Kerr nonlinearity in the linear dispersion model \cite{CalaLesina2016},
$P_i(t) = \chi^{(3)}_{iiii}E_i(t)|E(t)|^2$,
where $i=\{x,y,z\}$. 
We use $\chi^{(3)}_{Au}=7.6\cdot 10^{-19}$ $m^2/V^2$ \cite{Boyd} and neglect nonlinearity in the substrate.
The gap dielectric is a fictitious material having the linear properties of \chem{ITO} but a variable $\chi^{(3)}_{g}$ in order to explore the design space. We seek the value of $\chi^{(3)}_{g}$ for which the emission from the gap material and gold is the same. 
The simulation of the linear and nonlinear responses run at the same time, naturally taking into account pump depletion.
We simulate nine dipole nanoantennas (each combination of $w$ and $g$) and three monopole nanoantennas (each $w$ value), one at a time, and without periodic boundary conditions. 

To understand how the gap material and gold contribute to the overall nonlinear emission, we consider the following cases:
(I) emission only from the gap material in dipole nanoantennas ($\chi^{(3)}_{g}\neq 0$, $\chi^{(3)}_{Au} = 0$),
(II) emission only from gold in dipole nanoantennas ($\chi^{(3)}_{g}=0$, $\chi^{(3)}_{Au}\neq 0$), 
(III) emission from the gap material and gold in dipole nanoantennas ($\chi^{(3)}_{g}\neq 0$, $\chi^{(3)}_{Au}\neq 0$),
(IV) emission from gold in monopole nanoantennas ($\chi^{(3)}_{Au}\neq 0$).
We quantify the nonlinear emission by calculating the THG scattered power at $3\omega_0$ as the integral of the normal component of the Poynting vector through a closed surface in the far-field.
We report the relative THG efficiency $\eta_{THG}$ by normalizing with respect to the nanostructure with the largest THG emission, which is the monopole nanoantenna with $w=30$ nm.

\begin{figure}[htbp]
\centering
\includegraphics[width=0.5\textwidth]{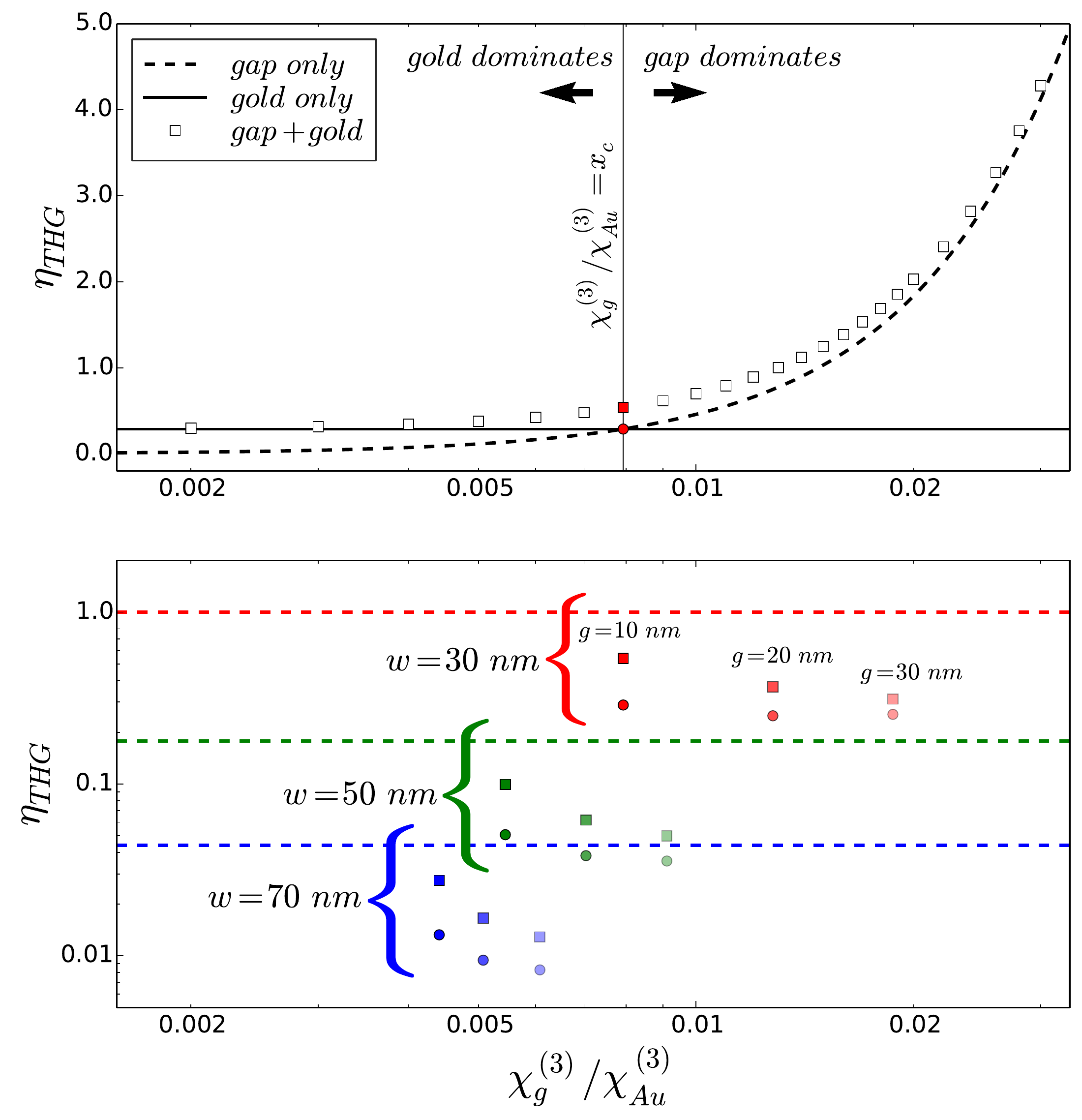}
\begin{picture}(0,0)
\put(-270,265){(a)}
\put(-270,132){(b)}
\end{picture}
\caption{(a) $\eta_{THG}$ for a dipole nanoantenna ($w=30$ nm, $g=10$ nm): emission from gap material only (dashed line), gold only (solid line), and both gap material and gold (squares). 
(b) $\eta_{THG}$ for the nine dipole nanoantennas ($\begingroup\color{black}\bullet\endgroup$,$\begingroup\color{black}\filledsquare\endgroup$) and the three monopole nanoantennas (dashed lines): $w=30$ nm (red), $w=50$ nm (green), and $w=70$ nm (blue), $g=10$ nm (dark shade), $g=20$ nm (medium shade), $g=30$ nm (light shade).}
\label{fig2}
\end{figure}

In Fig. \ref{fig2}a, we plot $\eta_{THG}$ vs $\chi^{(3)}_{g}/\chi^{(3)}_{Au}$ for case (I), (II) and (III) for the dipole nanoantenna ($w=30$ nm, $g=10$ nm) with the highest $\eta_{THG}$.
The black dashed line is for emission from the gap material only -- case study (I) -- and shows the expected quadratic behaviour with $\chi^{(3)}_{g}/\chi^{(3)}_{Au}$. The black line is for emission from gold only -- case study (II) -- which is horizontal since we are considering a single value for $\chi^{(3)}_{Au}$. The cross point ($\chi^{(3)}_{g}/\chi^{(3)}_{Au}=x_{c}$) between dotted and horizontal lines -- red circle ($\begingroup\color{red}\bullet\endgroup$) -- identifies the value such that the nonlinear emission from gold equals that from the gap material. The white squares are for emission from gold and gap -- case study (III) -- and they show an asymptotic behaviour, approaching the horizontal line on the left (gold only) and the quadratic curve on the right (gap only). The red square ($\begingroup\color{red}\filledsquare\endgroup$) identifies the case study (III) for $\chi^{(3)}_{g}=x_{c}\cdot\chi^{(3)}_{Au}$. For $\chi^{(3)}_{g}/\chi^{(3)}_{Au}>>x_{c}$ emission from the gap dominates, while for $\chi^{(3)}_{g}/\chi^{(3)}_{Au}<<x_{c}$ the emission from gold dominates.

\begin{figure}[htbp]
\centering
\includegraphics[width=0.52\textwidth]{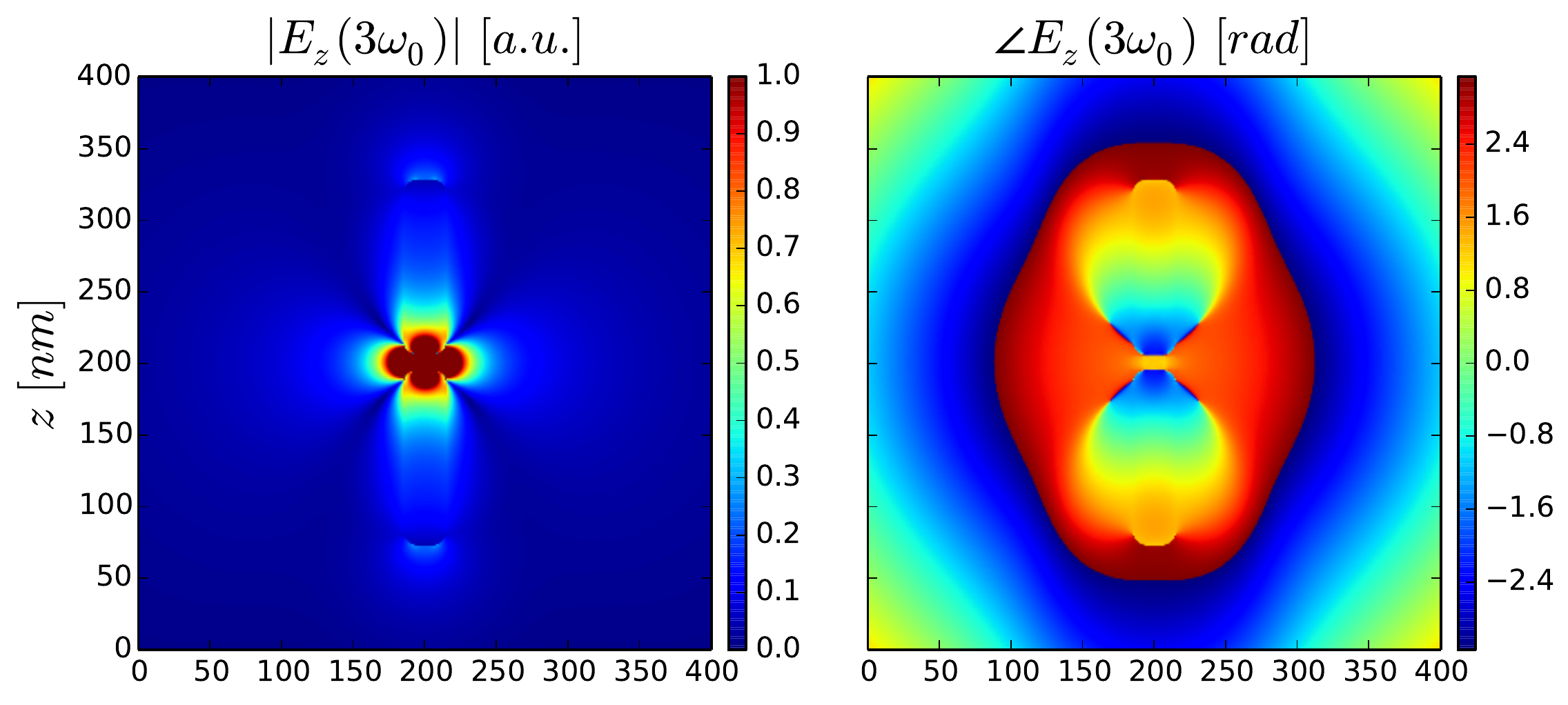}
\includegraphics[width=0.52\textwidth]{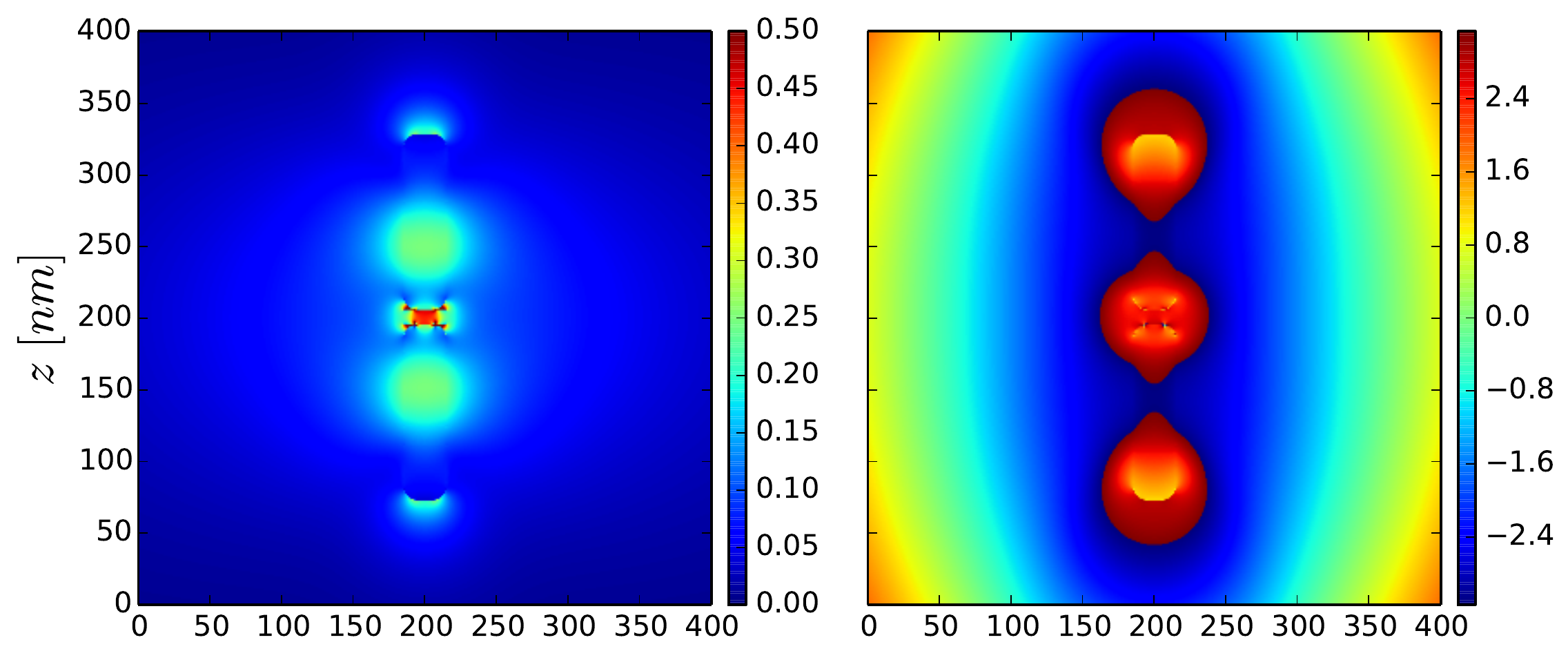}
\includegraphics[width=0.52\textwidth]{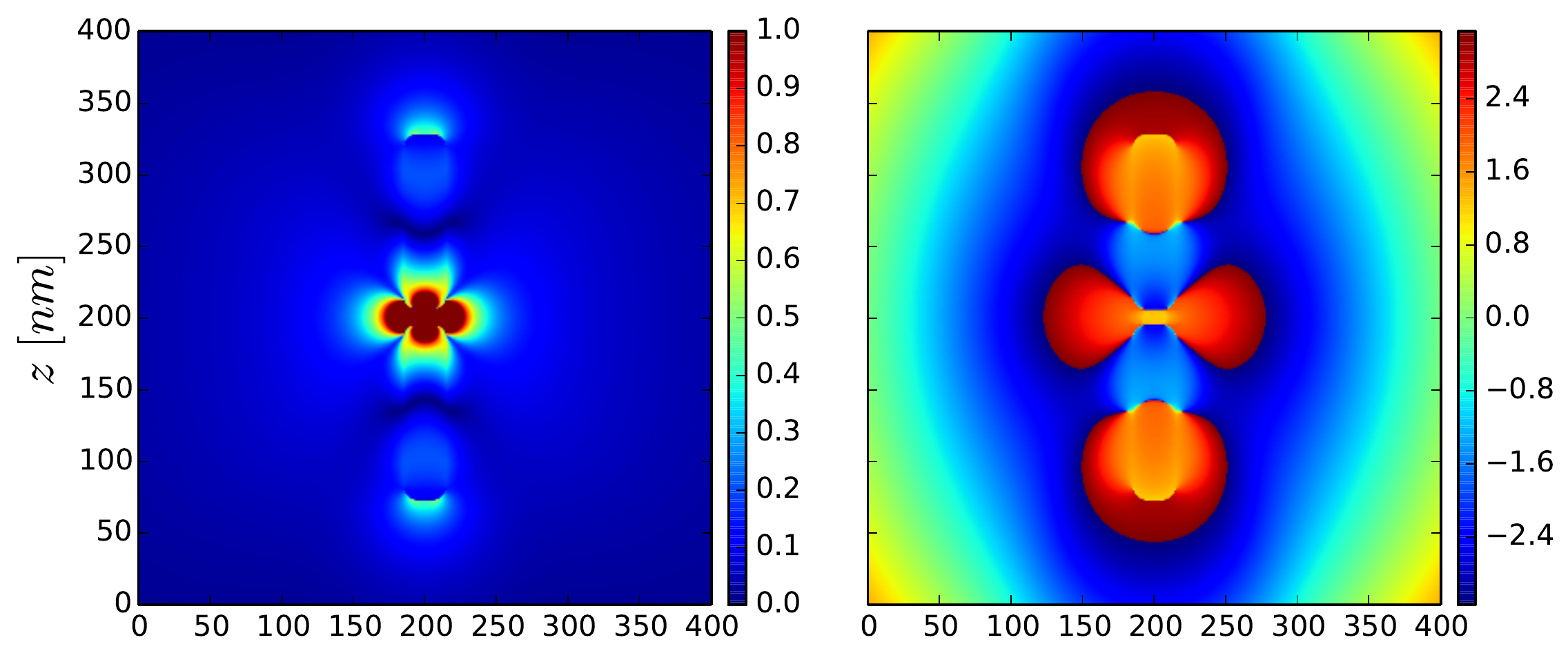}
\includegraphics[width=0.52\textwidth]{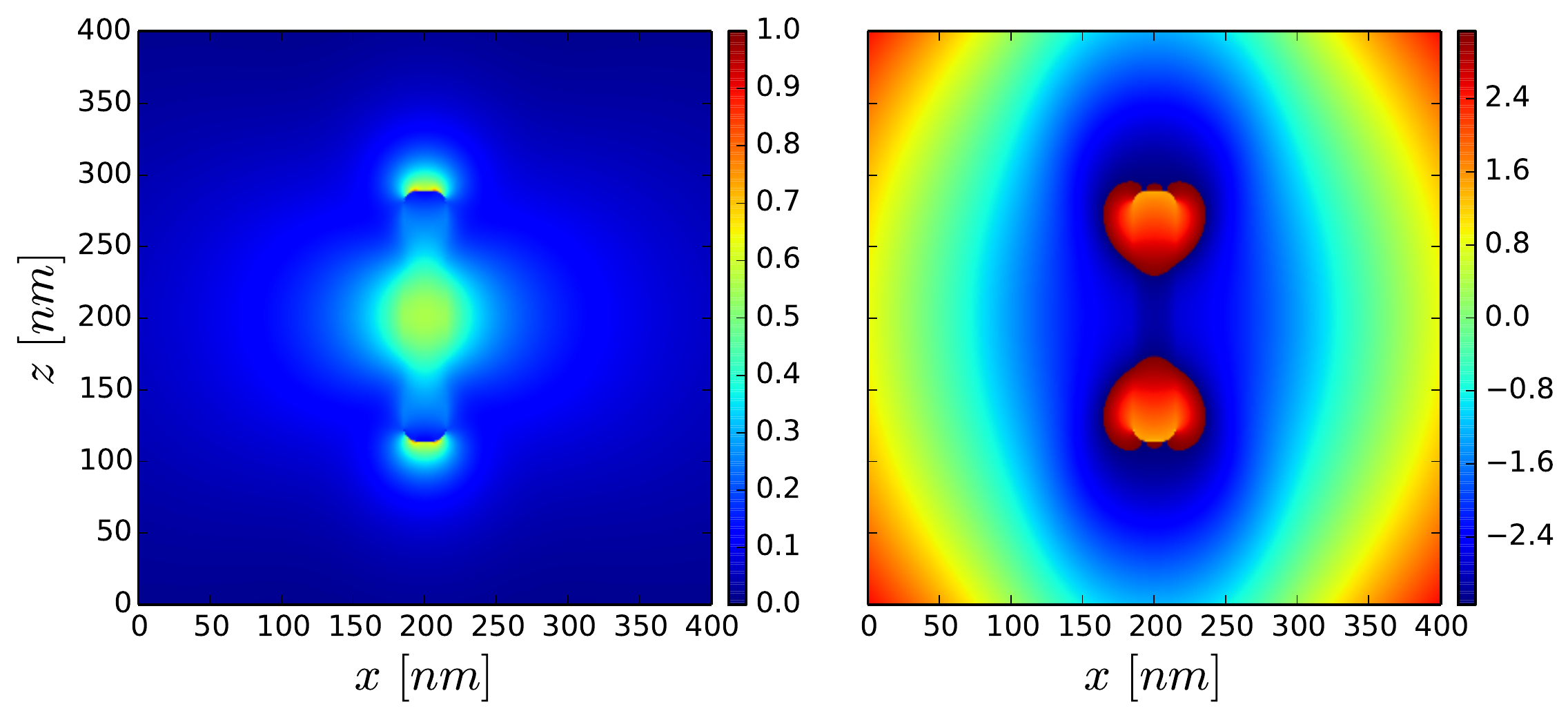}
\begin{picture}(0,0)
\put(-290,468){(a)}
\put(-290,352){(b)}
\put(-290,236){(c)}
\put(-290,120){(d)}
\end{picture}
\caption{Field distribution for $E_z(3\omega_0)$: emission from (a) gap material only, (b) gold only, and (c) gap material and gold in a dipole nanoantenna; (d) gold in a monopole nanoantenna.}
\label{fig3}
\end{figure}
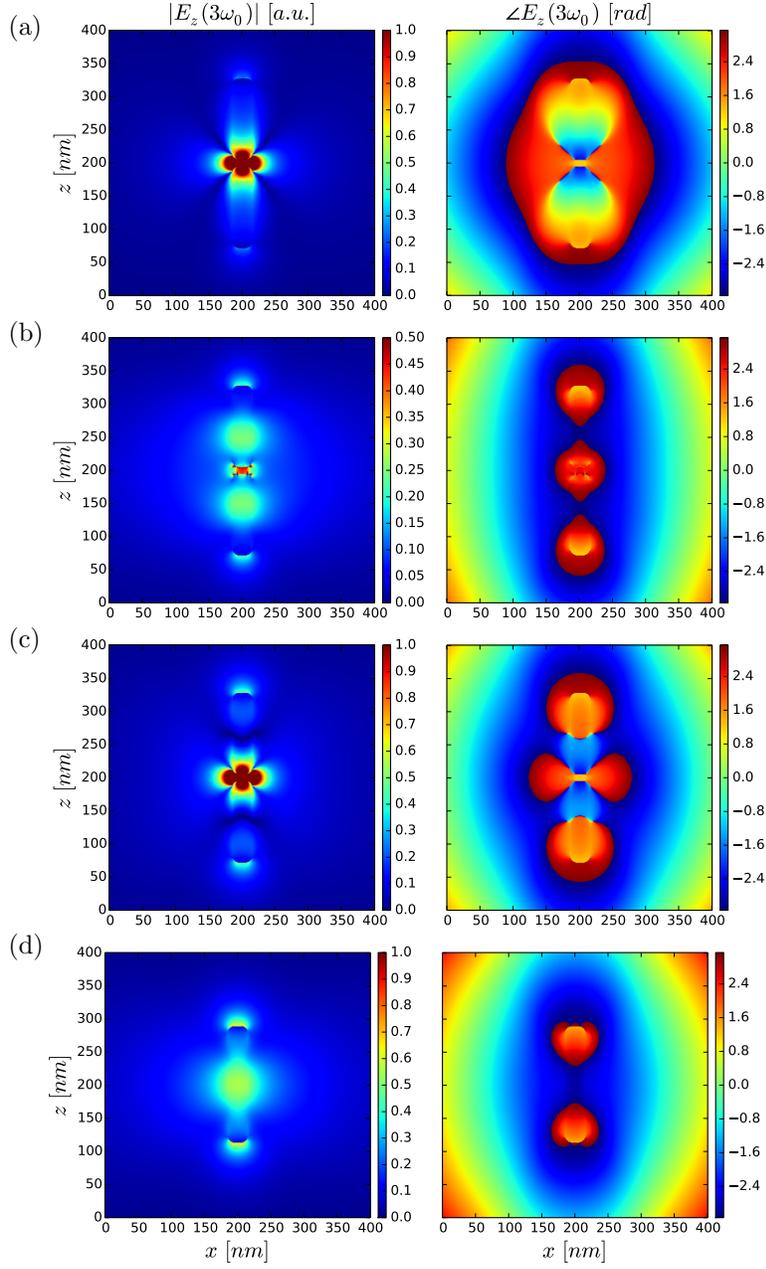

The red circle and square in Fig. \ref{fig2}a are sufficient to identify the boundary between gold and gap dominated regimes, as well as $\eta_{THG}$ at the cross point $x_c$. Thus, only these two points are shown in Fig. \ref{fig2}b for each of the nine dipole nanoantennas. For comparison, dotted lines are used to indicate $\eta_{THG}$ of the three monopoles -- case study (IV). 
For the dipole nanoantenna, by changing $w$ and $g$ we observe the variation of four parameters: $\eta_{THG}$, $x_c$, the horizontal spacing between circles ($d_h$), and the vertical spacing between circle and square ($d_v$).
Increasing $w$ produces a strong decrease in $\eta_{THG}$, a decrease in $x_c$ and a decrease in $d_h$.
Increasing $g$ produces a slight decrease in $\eta_{THG}$, an increase in $x_c$, and a decrease in $d_v$. 
A high $\eta_{THG}$ means that the baseline emission from gold is high. A high $x_{c}$ means that the emission from gold is more prone to dominate over the emission from the gap material. 
The lower $d_h$ for the largest width indicates that a larger gap volume compensates the lower field enhancement. The smaller $d_v$ for the largest gap size means that the gap material's contribution to $\eta_{THG}$ is small.   
The monopole nanoantenna always performs better than the dipole nanoantennas with the same $w$, since a continuous structure supports a less damped linear harmonic oscillation, which is a cause of strong nonlinear emission \cite{Hentschel2012}.

In the linear regime, the spectral peak of the near-field enhancement in the gap is slightly red-shifted with respect to the peak of the extinction spectrum \cite{Zuloaga2011}.
The nonlinear emission by the gap material is driven by the linear field enhancement in the gap, while the nonlinear emission from gold is linked to the linear extinction. 
Our simulations show that the nonlinear spectra for the emission from the gap and for the emission from gold exhibit a shift in wavelength, as also confirmed in \cite{DeCeglia2016}.
We use the linear extinction spectrum to determine the nanoantenna lengths required for resonance at $\lambda_0$. This corresponds to maximizing the nonlinear emission from gold. Thus, the nonlinear emission from the gap is not optimal. For example, in the case of the dipole nanoantenna with $w=30$ nm and $g=10$ nm, the linear extinction spectrum has its peak at $\sim 900$ nm, and its maximum linear near-field in the gap at $\sim 910$ nm ($\sim 10$ nm shift). The peak for the nonlinear emission from the gap was found at $\sim 305$ nm, while the peak for the nonlinear emission from gold was found at $\sim 300$ nm. Since we are considering the pump at 900 nm and the THG at 300 nm, we are underestimating the nonlinear emission from gap by $\sim 10\%$. 
Increasing $w$ and $g$ causes an increase in the shift between the spectral peak of the linear field enhancement in the gap and the peak of the linear extinction spectrum, up to $\sim 35$ nm for $w=70$ nm and $g=30$ nm.

In Fig. \ref{fig3}, we plot the near-fields for the dipole nanoantenna ($w=30$ nm, $g=10$ nm) and the monopole nanoantenna ($w=30$ nm) with highest $\eta_{THG}$.
We show the absolute value and phase of $E_z(3\omega_0)$ in the $xz$ plane for cases (I) through (IV) in Figs. \ref{fig3}a through \ref{fig3}d, respectively, assuming $\chi^{(3)}_{g}=x_{c}\cdot\chi^{(3)}_{Au}$. The electric field hot spots show nearly uniform phase and thus can be considered as idealized Huygens sources. 
In the case of THG emission from the gap, as expected, the Huygens source is localized in the gap (Fig. \ref{fig3}a).
More interesting is the emission from gold in the dipole nanoantenna, where we observe one Huygens source in each branch of the dipole nanoantenna, and an enhancement of the nonlinear field in the gap (Fig. \ref{fig3}b). The electric field distribution when both the gap and gold are emitting (Fig. \ref{fig3}c) shows constructive interference in the gap and at the extremes of the dipole nanoantanna, and destructive interference along the dipole nanoantenna branches. The monopole shows a Huygens source in the middle (Fig. \ref{fig3}d), as also reported in \cite{Wolf2016}. 
These nonlinear Huygens sources can be used in Huygens metasurfaces for nonlinear beam shaping \cite{CalaLesina2016}.

\begin{figure}[htbp]
\centering
\includegraphics[width=0.23\textwidth]{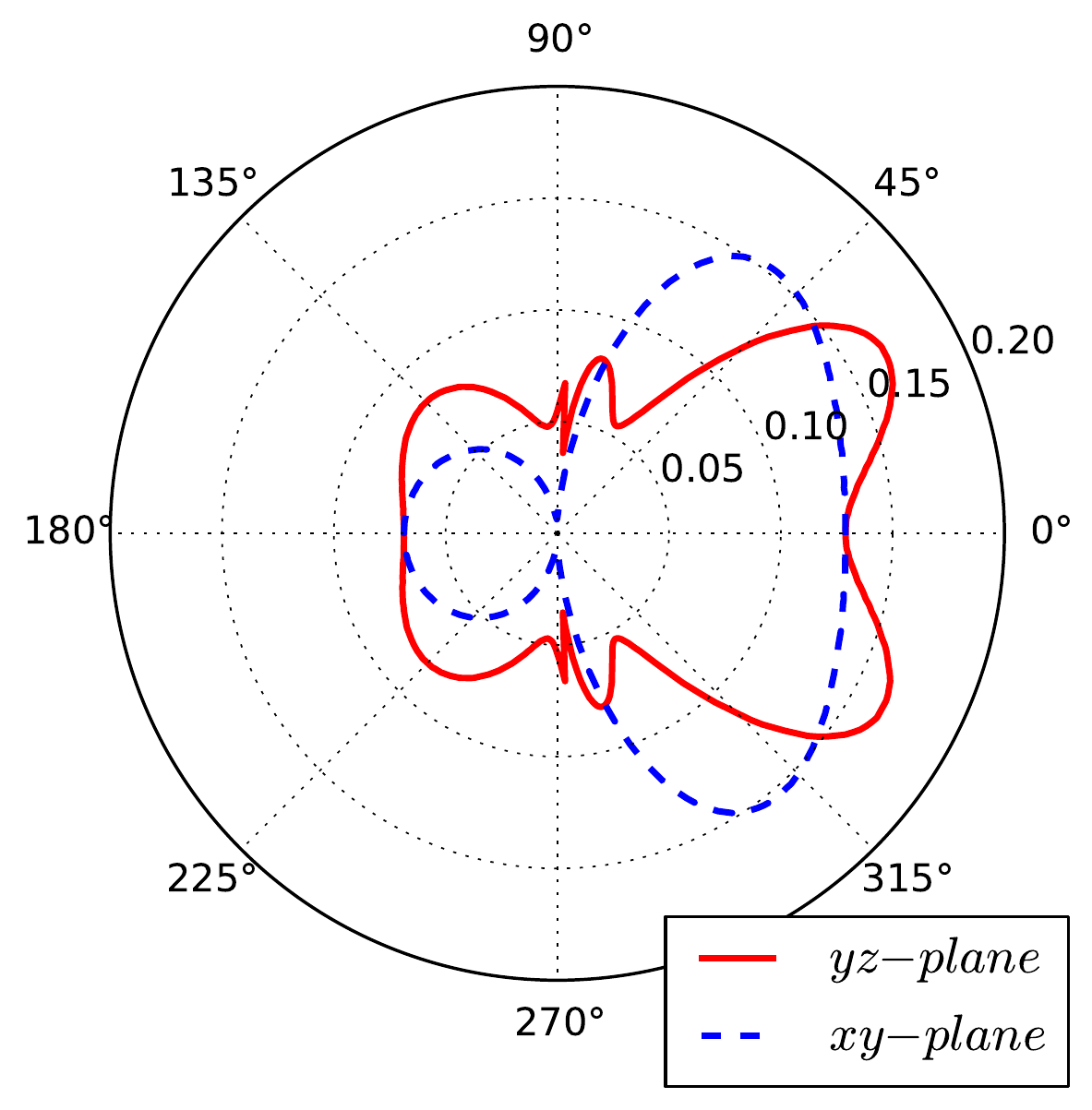}
\includegraphics[width=0.23\textwidth]{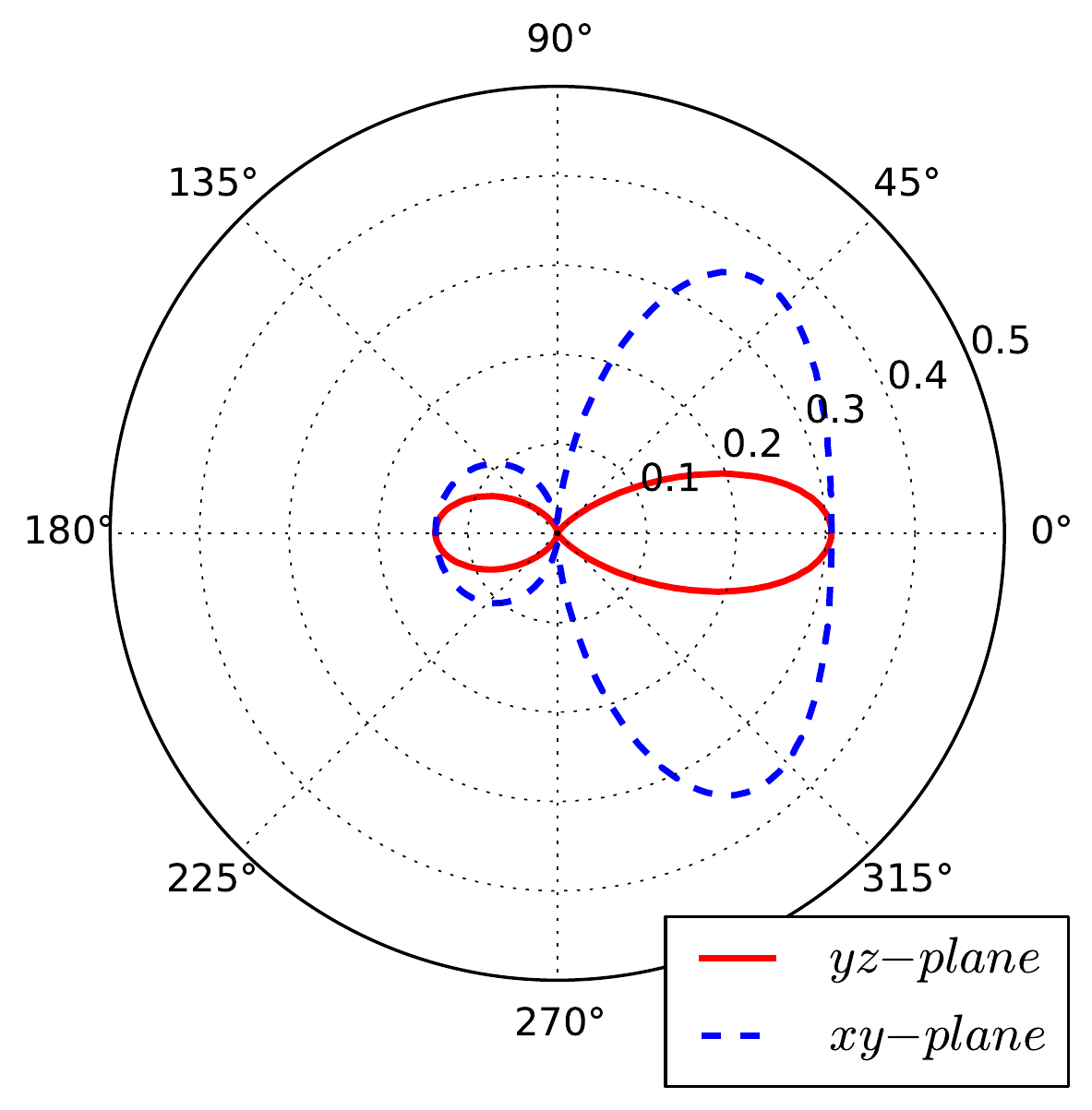}
\includegraphics[width=0.23\textwidth]{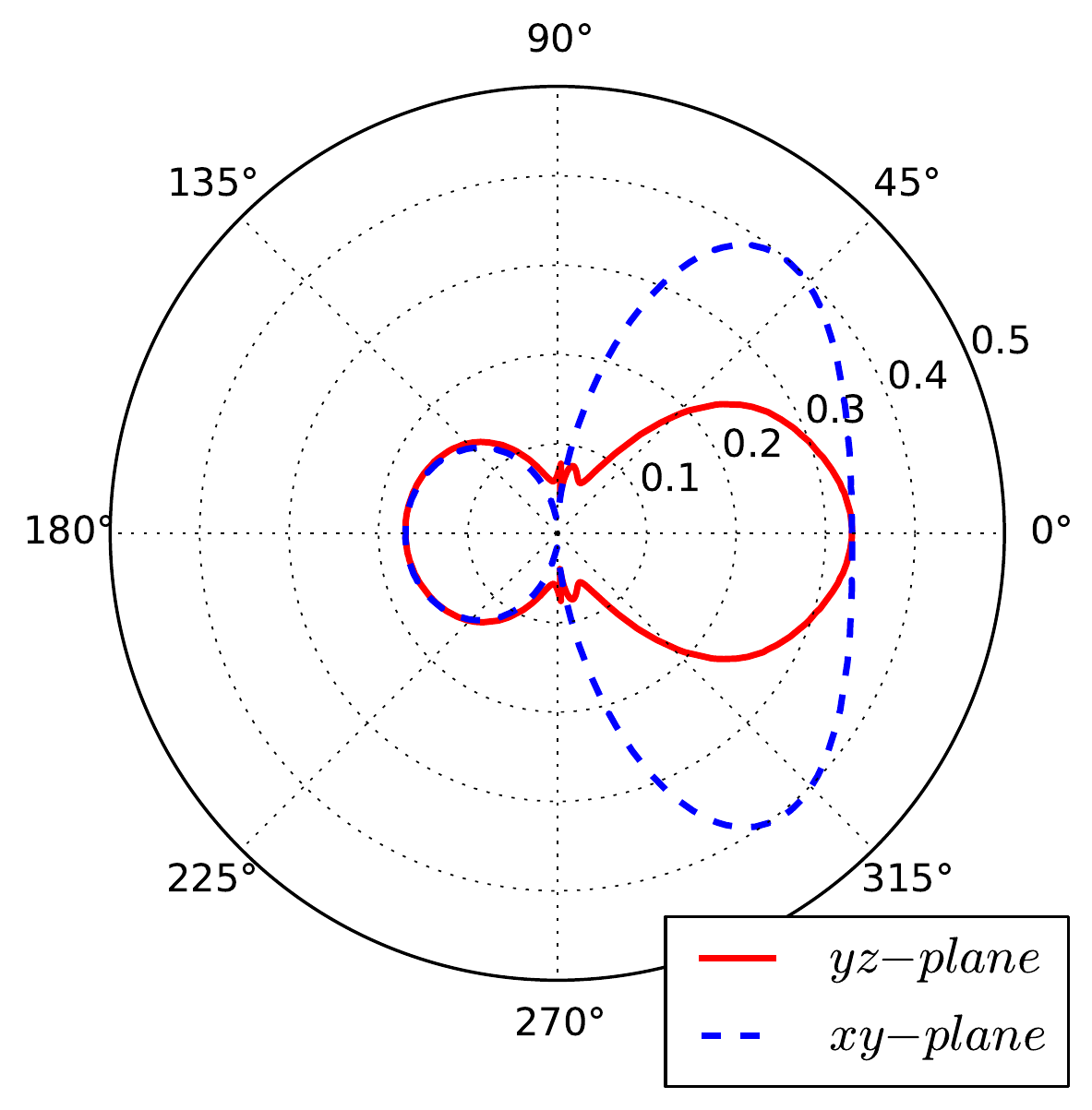}
\includegraphics[width=0.23\textwidth]{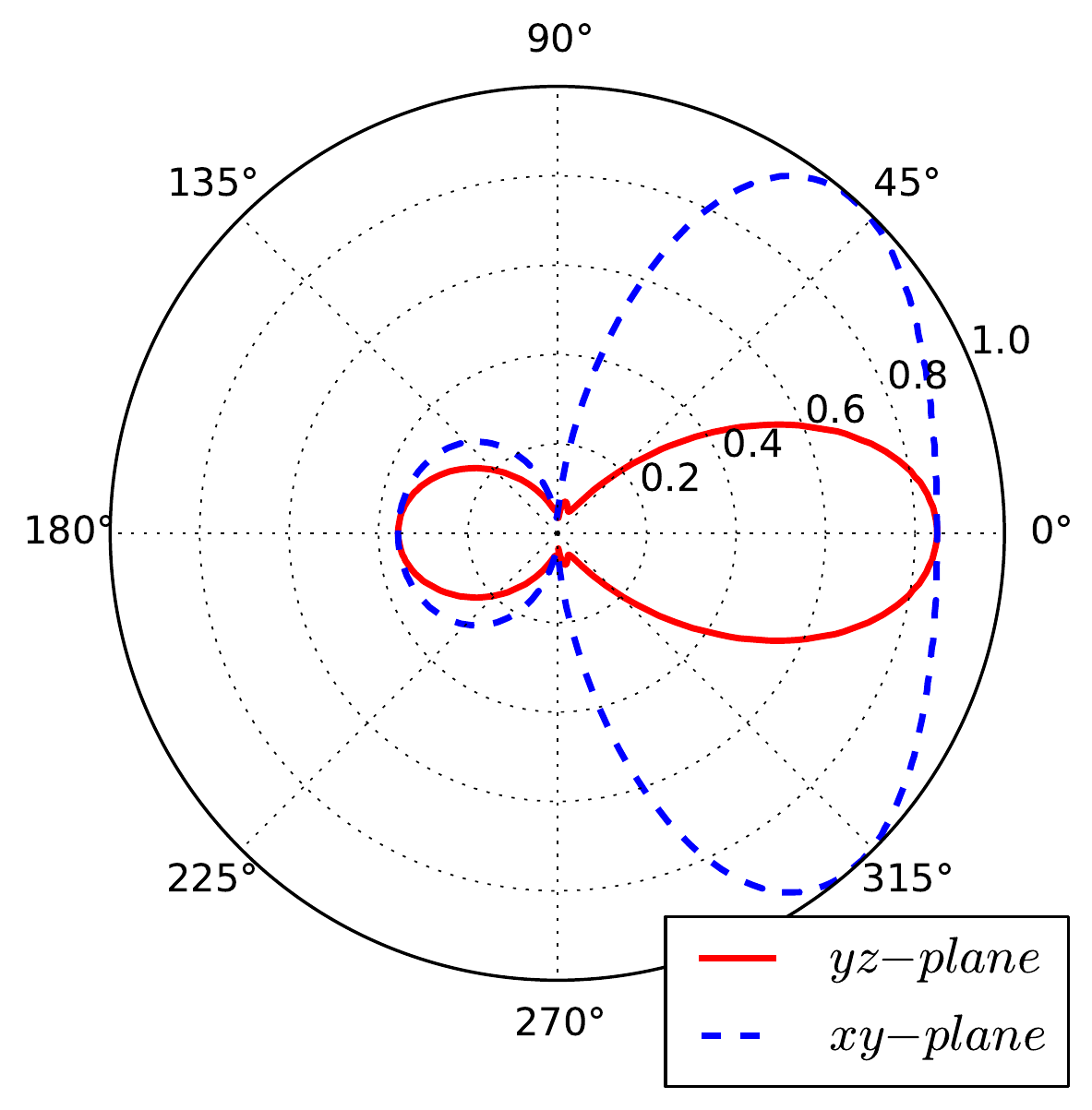}
\begin{picture}(0,0)
\put(-510,120){(a)}
\put(-380,120){(b)}
\put(-250,120){(c)}
\put(-120,120){(d)}
\end{picture}
\caption{Radiation pattern at $3\omega_0$: emission from (a) gap material only, (b) gold only, and (c) gap material and gold in a dipole nanoantenna; (d) gold in a monopole nanoantenna.}
\label{fig4}
\end{figure}
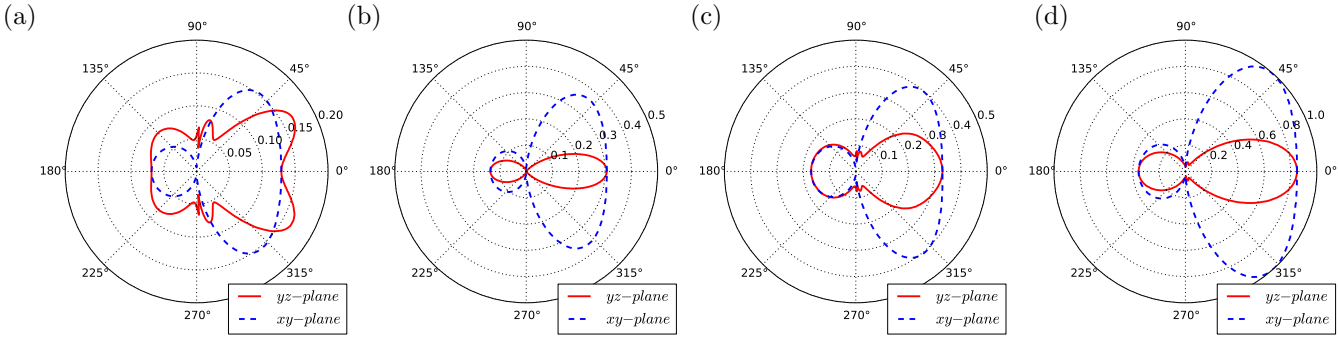

In Fig. \ref{fig4} we show the far-field radiation patterns at $3\omega_0$ corresponding to the nanoantennas in Fig. \ref{fig3}. 
We consider $xy$ and $yz$ cut planes, where the angle is measured starting from the $y$-axis. The radiation patterns were calculated by considering a polar coordinate system with the scattering centre in the middle of the nanoantenna, and evaluating the normal projection of the Poynting vector through a circle in the far-field. The radiation patterns are normalized with respect to the monopole with $w=30$ nm. 
The radiation can be seen as produced by a collinear array of phased emitters aligned along $z$ with subwavelength spacing. This produces radiation patterns in the $xy$ plane with the same shape in the four cases in Fig. \ref{fig4}. 
In all cases, the radiation is observed to be primarily directed through the substrate because it has a larger refractive index than the air above the antennas.
The different electric field distributions shown in Fig. \ref{fig3} play a role only for the radiation patterns in the $yz$ plane.
The emission from the gap material only produces a broad radiation pattern (Fig. \ref{fig4}a), which can be understood as diffraction by a small aperture; in fact, the broadening decreases as the gap size increases (not shown). 
In the case of emission from gold only in the dipole nanoantenna (Fig. \ref{fig4}b), the two nonlinear Huygens sources produce a directive lobe.
The simultaneous emission from gold and the gap material produces a hybrid radiation pattern (Fig. \ref{fig4}c).
The radiation pattern produced by the monopole nanoantenna (Fig. \ref{fig4}d) has the same shape as the one in Fig. \ref{fig4}b, but is less directive. This is because the central emitter dominates over the sources localized at the extremes of the monopole nanoantenna. Thus among the analyzed configurations, this is the structure which most closely resembles an ideal Huygens source. 
Furthermore, we have observed that the forward-to-backward ratio increases with $w$ when the emission comes from the gap, and decreases with $w$ when the emission comes from gold.
These radiation properties help to identify the position of the nonlinear emitters, and to understand which emission, whether from gap or gold, is dominating in an experiment.

As reported in \cite{Hentschel2012,Hanke2012}, nonlinear emission from metal nanostructures is larger for threadlike rather than bulky shapes. Our simulations showed that nonlinear emission from gold depends strongly on $w$ and much less on $g$. The main reason for the correct but seemingly conflicting conclusions in \cite{Metzger2014,Aouani2014} is found in the shape of the dipole nanoantennas and in the assumptions made about $\chi_g^{(3)}$. 
It is reasonable that the emission from gold dominates in \cite{Metzger2014} because the structure is threadlike ($w=50$ nm), and $P(3\omega)\propto P^3(\omega)$ is assumed, which may result in simulating a small $\chi_g^{(3)}$. However, the emission from the gap material dominates in \cite{Aouani2014} because the nanostructures are bulky (nanorod dimer with $w\sim 100$ nm and nanocylinder dimer with diameter $\sim 300$ nm) and a large $\chi_g^{(3)}$ is assumed.
Furthermore, the two papers consider different gap sizes ($g=20$ nm in \cite{Metzger2014} and $g>35$ nm in \cite{Aouani2014}), but as we demonstrated, the emission from gold is not strongly dependent on $g$.

\section{Conclusion}
In conclusion, we investigated third harmonic generation (THG) in hybrid dielectric/metal nanoantennas and found that both nanoantenna design and gap material nonlinear susceptibility determine whether the THG emission is primarily from the gap material, or from gold. In terms of design, nanoantenna width is a primary factor. In addition THG emission from the gold in a monopole is always much larger than that from the gold in a dipole of the same width. This is due to the lower damping of the linear harmonic oscillator which drives the nonlinear emission. 
Due to its fabrication simplicity, the monopole nanoantenna should be preferred, unless the nonlinear permittivity of the gap material is strong enough to make its nonlinear emission much larger than that from gold.

The authors would like to thank the Southern Ontario Smart Computing Innovation Platform (SOSCIP), SciNet, and the Canada Research Chairs program.

% Bibliography
\bibliography{Hybrid_nanoantenna}

%\end{multicols}
\end{document}